\documentclass{ws-ijmpcs}

\newcommand{\as}{\alpha_s}
\def\eq#1{{Eq.~(\ref{#1})}}
\def\fig#1{{Fig.~\ref{#1}}}

\begin{document}

\markboth{Yuri~V.~Kovchegov, Matthew D. Sievert}
{Single Spin Asymmetry in High Energy QCD}

%
\catchline{}{}{}{}{}
%

\title{SINGLE SPIN ASYMMETRY IN HIGH ENERGY QCD}

\author{YURI~V.~KOVCHEGOV \footnote{Presented by YK at the {\sl QCD
      Evolution Workshop}, May 14 - 17, 2012, Thomas Jefferson
    National Accelerator Facility, Newport News, VA.}}

\address{Department of Physics, The Ohio State University \\ Columbus, OH 43210, USA \\
kovchegov.1@asc.ohio-state.edu}

\author{MATTHEW D. SIEVERT}

\address{Department of Physics, The Ohio State University \\ Columbus, OH 43210, USA \\
sievert.7@osu.edu}

\maketitle


\begin{abstract}
  We present the first steps in an effort to incorporate the physics
  of transverse spin asymmetries into the saturation formalism of high
  energy QCD. We consider a simple model in which a transversely
  polarized quark scatters on a proton or nuclear target. Using the
  light-cone perturbation theory the hadron production cross section
  can be written as a convolution of the light-cone wave function
  squared and the interaction with the target. To generate the single
  transverse spin asymmetry (STSA) either the wave function squared or
  the interaction with the target has to be $T$-odd. In this work we
  use the lowest-order $q \to q \, G$ wave function squared, which is
  $T$-even, generating the STSA from the $T$-odd interaction with the
  target mediated by an odderon exchange. We study the properties of
  the obtained STSA, some of which are in qualitative agreement with
  experiment: STSA increases with increasing projectile $x_F$ and is a
  non-monotonic function of the transverse momentum $k_T$. Our
  mechanism predicts that the quark STSA in polarized proton--nucleus
  collisions should be much smaller than in polarized proton--proton
  collisions.  We also observe that the STSA for prompt photons due to
  our mechanism is zero within the accuracy of the approximation.

\end{abstract}

\ccode{PACS numbers: 24.85.+p, 12.38.Bx, 13.88.+e, 24.70.+s}


\section{Introduction}

This presentation is based on the paper [\refcite{Kovchegov:2012ga}].

Our goal is to use the saturation/Color Glass Condensate (CGC)
formalism (see
[\refcite{Iancu:2003xm,Weigert:2005us,Jalilian-Marian:2005jf}] for
reviews) to calculate the single transverse spin asymmetry (STSA) in
polarized proton--proton and polarized proton--nucleus collisions. The
observable is defined by
\begin{equation}
 \label{eq-Defn STSA}
   A_N ({\bm k}) \equiv \; \frac{\frac{d \sigma^\uparrow}{d^2 k \, dy} - 
   \frac{d \sigma^\downarrow}{d^2 k \, dy}} {\frac{d \sigma^\uparrow}{d^2 k \, dy} + 
   \frac{d \sigma^\downarrow}{d^2 k \, dy}} \; = \;
   \frac{\frac{d \sigma^\uparrow}{d^2 k \, dy} (\bm k) - 
   \frac{d \sigma^\uparrow}{d^2 k \, dy} (- \bm k)}
   {\frac{d \sigma^\uparrow}{d^2 k \, dy} (\bm k) + 
   \frac{d \sigma^\uparrow}{d^2 k \, dy} (- \bm k)}
   \; \equiv \frac{d(\Delta \sigma)}{2 \, d\sigma_{unp}}
\end{equation}
where the arrow indicates spin-up and spin-down outgoing hadron with
the spin direction taken here to be along the $\hat x$-axis (with
$\hat z$ the collision axis).

For simplicity we will consider scattering of a transversely polarized
quark on an unpolarized proton or nuclear target $q^{\uparrow} + A
\rightarrow q + X$. The realistic case of the incoming transversely
polarized proton can be recovered if one convolutes the cross section
we obtain with the polarized proton wave function squared, or, perhaps
equivalently, with the proton transversity distribution.

\begin{figure}[ht]
\begin{center}
 \includegraphics[width=0.45 \textwidth]{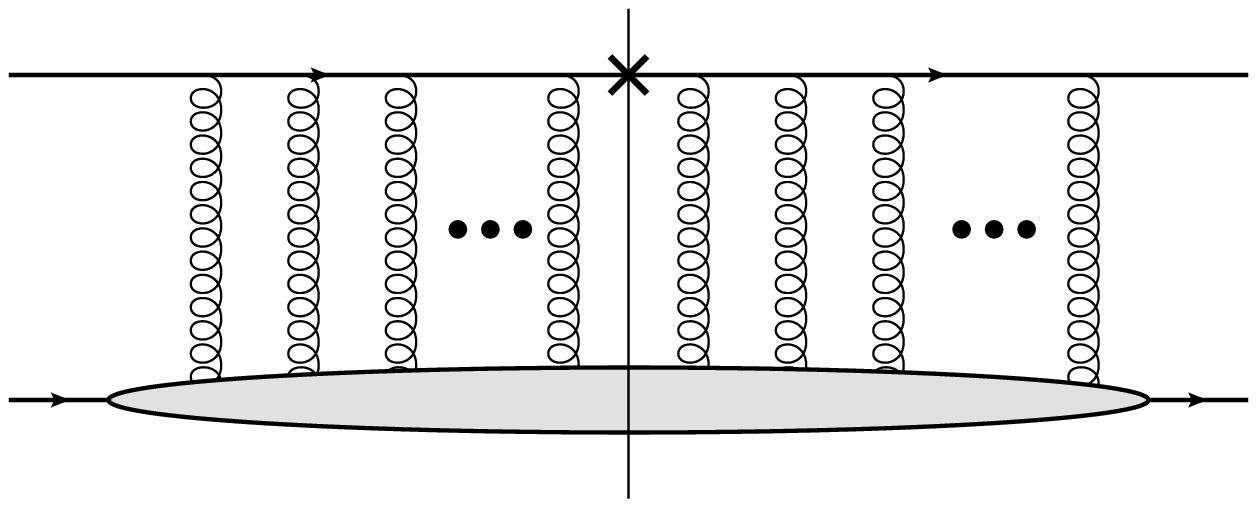} \ \ \ \includegraphics[width=0.45 \textwidth]{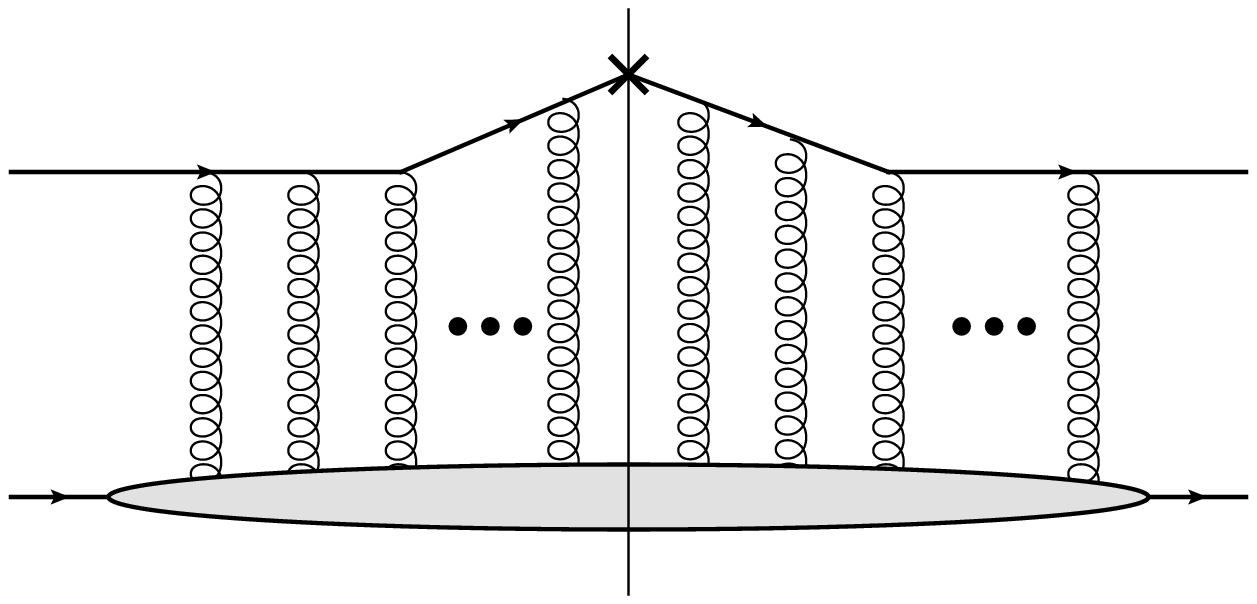} 
\end{center}
\caption{Left panel: eikonal scattering of a quark on a nucleus. Right
  panel: quark--nucleus scattering with recoil.}
\label{recoil} 
\end{figure}

Figure~\ref{recoil} demonstrates that to capture the spin-dependent
effects, such as the STSA, one has to go beyond the eikonal accuracy
in the calculation. Recoilless eikonal scattering, as shown in the
left panel of \fig{recoil}, is spin-independent and can not generate
the asymmetry. Naive inclusion of sub-eikonal corrections is possible
by introducing recoil in the scattering, as shown in the right panel
of \fig{recoil}: such diagrams are suppressed by a power of the center
of mass energy squared $s$ of the collision and are usually discarded
in the saturation/CGC formalism since they are very small.

A much larger contribution to STSA may arise from the non-eikonal
splitting of the projectile quark into a quark and a gluon, $q \to q
\, G$, which is suppressed only by a power of the strong coupling
$\as$. In the LCPT language the $q \to q \, G$ splitting may take
place either long before or long after the interaction with the
target, as shown in \fig{qtoqGampl} below. (Splitting during the
interaction with the target is suppressed by powers of energy
[\refcite{Kovchegov:1998bi}].) Both such contributions would be
included in the light-cone wave function squared of the projectile
(quark). 

The spin-dependence by itself is insufficient to generate the STSA:
one also needs a relative phase between the amplitude and the complex
conjugate amplitude contributing to the STSA-generating part of the
process [\refcite{Qiu:1998ia,Brodsky:2002cx}]. In the standard
interpretation of the Sivers effect [\refcite{Sivers:1989cc}] both the
spin-dependence and the relative phase originate in the wave function
squared (though, often-times, the actual diagrammatic representation
of the effect requires a final-state interaction with the polarized
projectile [\refcite{Brodsky:2002cx}], which, in the CGC
rapidity-ordered language, can still be incorporated as a final-state
light-cone wave function [\refcite{KS2}]). The lowest-order $q \to q
\, G$ splitting from \fig{qtoqGampl} is too basic to generate the
phase.  In the Collins mechanism the phase is generated in the
fragmentation function [\refcite{Collins:1992kk}]: this can certainly
take place in the saturation/CGC framework as well, but is not the
main aim of this investigation.

Below we show that it is also possible to generate the relative phase
in the integration with the unpolarized target. We thus obtain a new
mechanism for generating STSA, where the spin dependence comes from
the polarized projectile wave function, while the phase arises in the
interaction with the target.


\section{Calculation of STSA in the Saturation Framework}

\begin{figure}[h]
\includegraphics[width=\textwidth]{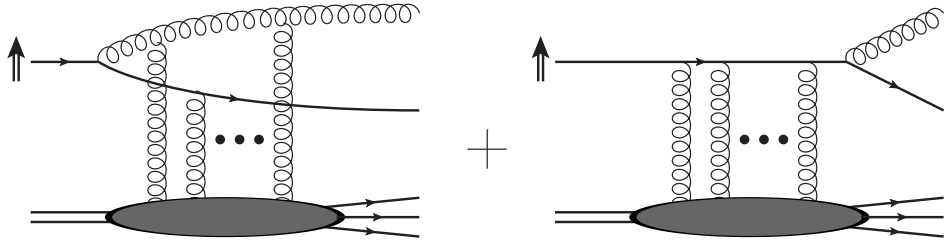}
 \caption{Two contributions to the amplitude for the high energy 
   quark--target scattering in LCPT.}
\label{qtoqGampl}  
\end{figure}

\begin{figure}[th]
 \includegraphics[width=\textwidth]{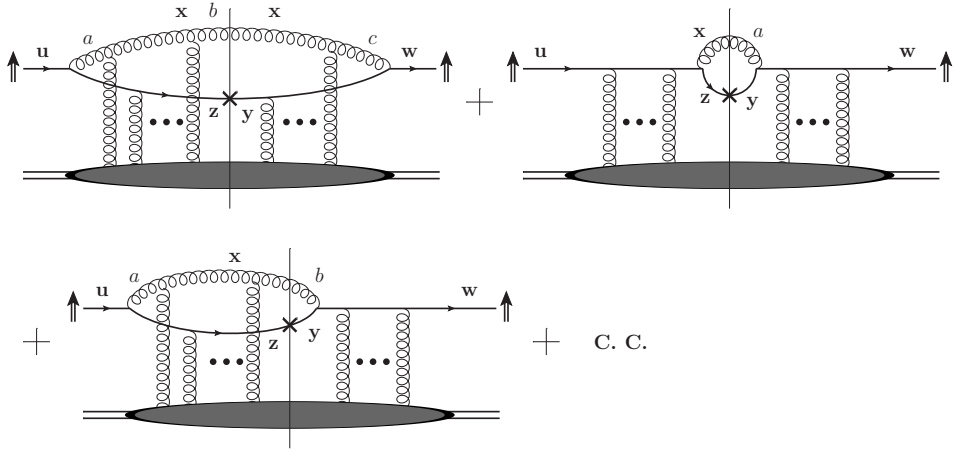}
 \caption{The cross section for quark production in the polarized quark--nucleus scattering.}
\label{fig-amplitude squared} 
\end{figure}

The diagrams for the scattering process of a transversely polarized
quark on a proton or nuclear target are shown in
\fig{qtoqGampl}. Squaring the sum of these diagrams gives us the
graphs shown in \fig{fig-amplitude squared} which contribute to the
inclusive quark production cross section.

The calculation proceeds along the standard steps employing the
light-cone perturbation theory (LCPT) (see
[\refcite{Jalilian-Marian:2005jf}] for a review): the $q \to q \, G$
splitting contributes to the light-cone wave function, which
factorizes from the interaction with the target. The general result
for the quark production in the $q^\uparrow + A$ scattering given by
diagrams in \fig{fig-amplitude squared} reads
\begin{equation}
 \label{eq-cross section}
 \frac{d\sigma^{(q)}}{d^2 k \, d y_q} = \frac{C_F}{2 \, (2\pi)^3} \, 
 \frac{\alpha}{1-\alpha} \, \int
 d^2 x \, d^2 y \, d^2 z \, e^{-i \bm k \cdot (\bm z - \bm y)} \,
 \Phi_\chi (\bm z - \bm x \, , \, \bm y - \bm x, \alpha) \
 \mathcal{I}^{(q)} (\bm x \, , \, \bm y \, , \, \bm z)
\end{equation}
with $\Phi_\chi$ denoting the light-cone wave function squared and
$\mathcal{I}^{(q)}$ representing the interaction with the target with
all the coordinate labels shown explicitly in \fig{fig-amplitude
  squared}. Here
%
%
\begin{subequations}
 \label{eq-kinematic constraints}
 \begin{eqnarray}
  \bm u = \bm x + \alpha \, (\bm z - \bm x) \\ 
  \bm w = \bm x + \alpha \, (\bm y - \bm x)
 \end{eqnarray}
\end{subequations}
are the transverse coordinates of the incoming quark in the amplitude
and in the complex conjugate amplitude, while $\alpha$ is the fraction
of the incoming quark's plus momentum component $p^+$ carried by the
produced quark with the plus momentum component $k^+$, such that
$\alpha = k^+/p^+$.

The light-cone wave function squared $\Phi_\chi$ can be decomposed
into a polarization-independent term and another term which is
proportional to the quark transverse polarization eigenvalue $\chi$,
\begin{eqnarray}
 \label{eq-wavefn squared}
 \Phi_\chi (\bm z - \bm x , \bm y - \bm x, \alpha) 
 = \, \Phi_{unp} (\bm z - \bm x , \bm y 
 - \bm x, \alpha) + \chi \, \Phi_{pol} (\bm z - \bm x , \bm y - \bm x, \alpha) 
 \; ,
\end{eqnarray}
where a straightforward calculation yields [\refcite{Kovchegov:2012ga}]
\begin{eqnarray}
 \label{eq-unpolarized wavefn}
 \Phi_{unp} &=& \frac{2 \, \alpha_s}{\pi} \, {\tilde m}^2 \, \bigg[ (1+\alpha^2) \, \frac{(\bm z -   \bm x) \cdot (\bm y - \bm x)}{|\bm z - \bm x| \;       
 |\bm  y - \bm x|} \, K_1 (\tilde m \, |\bm z - \bm x|) \, K_1 (\tilde m \, 
 |\bm y - \bm x|) \\ \nonumber
 && + \, (1-\alpha)^2 \, K_0 (\tilde m \, |\bm z - \bm x|) \, K_0 (\tilde m \,  
 |\bm y - \bm x|) \bigg]
\end{eqnarray}
for the unpolarized component and
\begin{eqnarray}
 \label{eq-polarized wavefn}
 \Phi_{pol} = \frac{2 \, \alpha_s}{\pi} \, {\tilde m}^2 \, \alpha \, (1-\alpha) \, \bigg[
 \frac{z^{2} - x^{2}}{|\bm z - \bm x|} \, K_0 (\tilde m \, |\bm y - 
 \bm x|) \, K_1 (\tilde m \, |\bm z - \bm x|) \notag \\
  + \, \frac{y^{2} - x^{2}}{|\bm y - \bm x|} \, K_1 (\tilde m \, |\bm y -
 \bm x|) \, K_0 (\tilde m \, |\bm z - \bm x|) \bigg] 
\end{eqnarray}
for the polarization-dependent part. Here ${\tilde m} = m \,
(1-\alpha)$ with $m$ the quark mass.

To describe the interaction with the target in \fig{fig-amplitude
  squared} we use the Wilson line formalism in the $A^+ =0$ light cone
gauge of the projectile (see [\refcite{Weigert:2005us}] for a detailed
description of the formalism). (Our light cone 4-vector components are
defined by $x^\pm = t\pm z$.) Defining the fundamental representation
Wilson line
\begin{equation}
  \label{eq:Wline}
  V_{\bm x} \equiv {\mathcal P} \exp \left[ \frac{i \, g}{2} \, 
    \int\limits_{-\infty}^{+\infty} d x^+ \, T^a \, A^{- \, a} 
    (x^+, x^- =0, \bm x ) \right]
\end{equation}
we can write the $S$-matrix operator for a fundamental-representation
color dipole scattering on the target proton or nucleus by
\begin{equation}
  \label{Ddef}
  {\hat D}_{\bm x \, \bm y} \equiv \frac{1}{N_c} \, \mathrm{Tr} \, \left[ V_{\bm x}
 \, V^\dagger_{\bm y} \right].
\end{equation}
Using this object we can write the interaction with the target
$\mathcal{I}^{(q)}$ as
\begin{eqnarray}
  \label{Iq}
  \mathcal{I}^{(q)} = \left\langle {\hat D}_{\bm z \, \bm y} + {\hat D}_{\bm u \, \bm w} 
    - \frac{N_c}{2 \, C_F} \, {\hat D}_{\bm z \, \bm x} \, {\hat D}_{\bm x \, \bm w} + 
    \frac{1}{2 \, N_c \, C_F} \, {\hat D}_{\bm z \, \bm w} \right. \\ \notag \left. - 
    \frac{N_c}{2 \, C_F} \, {\hat D}_{\bm u \, \bm x} \, {\hat D}_{\bm x \, \bm y} + 
    \frac{1}{2 \, N_c \, C_F} \, {\hat D}_{\bm u \, \bm y} \right\rangle ,
\end{eqnarray}
where the first two terms are given by the graphs in the first line of
\fig{fig-amplitude squared}.

Our goal is to find whether the quark production cross section
\eqref{eq-cross section} contributes to the STSA defined in
\eq{eq-Defn STSA}. Clearly STSA should be given by the part of the
cross section odd under the $\bm k \rightarrow - \bm k$
interchange. STSA is usually thought of as arising from the $({\vec S}
\times {\vec p} ) \cdot {\vec k}$ term in the cross section, where
$\vec S$ is the spin of the incoming projectile and $\vec p$ is its
momentum: such term is $T$-odd. Analyzing \eq{eq-cross section} we see
that the $\bm k \rightarrow - \bm k$ transformation can be thought of
as the $\bm z \leftrightarrow \bm y$ coordinate interchange. The
$T$-odd STSA-generating part results from the part of the integrand in
\eq{eq-cross section} which is odd under the $\bm z \leftrightarrow
\bm y$ interchange: since the integrand is a product of $\Phi_\chi$
and $\mathcal{I}^{(q)}$ the $T$-odd contribution may arise in either
of these terms. Because our lowest-order $\Phi_\chi$ is symmetric
under $\bm z \leftrightarrow \bm y$, at this level of approximation
the STSA can only arise in the interaction with the target
$\mathcal{I}^{(q)}$. Therefore we need to decompose \eq{Iq} into the
$\bm z \leftrightarrow \bm y$ symmetric and anti-symmetric parts, with
the latter one giving rise to STSA when used in \eq{eq-cross section}.

Before we proceed to anti-symmetrize $\mathcal{I}^{(q)}$, let us pause
and point out that the decomposition into the light cone wave function
squared and the interaction with the target is quite general and is
valid up to corrections suppressed by powers of energy, which are
usually neglected in the high energy QCD framework. The conclusion
that the $T$-odd STSA-generating term may arise either in the wave
function squared or in the interaction with the target is also quite
general. An example of the wave function squared giving the
STSA-generating contribution is the initial-state Sivers effect
[\refcite{Sivers:1989cc}], though, as was illustrated in
[\refcite{Brodsky:2002cx}], and as is also seen in \fig{qtoqGampl},
the wave function squared also includes final state splittings and
interactions. Below we will construct the first ever example of the
$T$-odd STSA-generating term generated in the interaction with the
target.

To anti-symmetrize $\mathcal{I}^{(q)}$ under the $\bm z
\leftrightarrow \bm y$ interchange we first need to decompose each
dipole $S$-matrix into the even and odd pieces under the exchange of
its transverse coordinates, which corresponds to the $C$-parity
operation exchanging the quark and the anti-quark
[\refcite{Kovchegov:2003dm,Hatta:2005as}]:
\begin{subequations}
 \label{eq-Pomeron and Odderon}
 \begin{eqnarray}
  {\hat D}_{\bm x \, \bm y} &\equiv& {\hat S}_{\bm x \, \bm y} + i \,
  {\hat O}_{\bm x \, \bm y} \\
  {\hat S}_{\bm x \, \bm y} &\equiv& \frac{1}{2} \, ( {\hat D}_{\bm x \, \bm y} +
  {\hat D}_{\bm y \, \bm x}) \label{Sdef} \\
  {\hat O}_{\bm x \, \bm y} &\equiv& \frac{1}{2i} \, ( {\hat D}_{\bm x \, \bm y} -
  {\hat D}_{\bm y \, \bm x}) \; . \label{Odef}
 \end{eqnarray}
\end{subequations}
The $C$- and $T$-even part ${\hat S}_{\bm x \, \bm y}$ is the
``standard'' dipole amplitude giving the total unpolarized DIS cross
section in DIS. Its expression in terms of Wilson lines allows to
include Glauber-Mueller multiple rescatterings
[\refcite{Mueller:1989st}] along with the nonlinear small-$x$
BK/JIMWLK evolution in the leading-$\ln s$ approximation. The $C$- and
$T$-odd {\sl odderon} amplitude ${\hat O}_{\bm x \, \bm y}$
[\refcite{Lukaszuk:1973nt,Nicolescu:1990ii,Ewerz:2003xi}] starts out
with the triple gluon exchange at the lowest order
[\refcite{Jeon:2005cf}], but then can also be enhanced by the multiple
rescatterings and small-$x$ evolution
[\refcite{Bartels:1999yt,Kovchegov:2003dm,Hatta:2005as,Kovner:2005qj}].

Using the decomposition \eqref{eq-Pomeron and Odderon} we can split
$\mathcal{I}^{(q)}$ into the $T$- and $C$-even $\bm z \leftrightarrow
\bm y$ symmetric part and the $T$- and $C$-odd $\bm z \leftrightarrow
\bm y$ anti-symmetric part:
\begin{subequations}
 \label{eq-symmetrized interaction}
 \begin{eqnarray}
  \mathcal{I}_{symm}^{(q)} &=& \left\langle {\hat S}_{\bm z \, \bm y} +
  {\hat S}_{\bm u \, \bm w} - \frac{N_c}{2 \, C_F} \, \left( {\hat S}_{\bm z \, \bm x} \, 
  {\hat S}_{\bm x \, \bm w} - {\hat O}_{\bm z \,  \bm x} \, {\hat O}_{\bm x \, \bm w} \right) 
  + \frac{1}{2 \, N_c \, C_F} \, {\hat S}_{\bm z \, \bm w} \right. \notag \\ 
  && \left. - \frac{N_c}{2 \, C_F} \, \left(
  {\hat S}_{\bm u \, \bm x} \, {\hat S}_{\bm x \, \bm y} -  {\hat O}_{\bm u \, \bm x} \, 
  {\hat O}_{\bm x \, \bm y} \right) 
  + \frac{1}{2 \, N_c \, C_F} \, {\hat S}_{\bm u \, \bm y} \right\rangle , \\
  \mathcal{I}_{anti}^{(q)} &=& i \, \left\langle {\hat O}_{\bm z \, \bm y} +
  {\hat O}_{\bm u \, \bm w} - \frac{N_c}{2 \, C_F} \, 
  \left( {\hat O}_{\bm z \, \bm x} \, {\hat S}_{\bm x \, \bm w} 
   + {\hat S}_{\bm z \, \bm x} \, {\hat O}_{\bm x \, \bm w} \right) 
  + \frac{1}{2 \, N_c \, C_F} \, {\hat O}_{\bm z \, \bm w} \right. \notag \\ 
  && \left. - \frac{N_c}{2 \, C_F} \, 
  \left(
  {\hat O}_{\bm u \, \bm x} \, {\hat S}_{\bm x \, \bm y} +  {\hat S}_{\bm u \, \bm x} \,
  {\hat O}_{\bm x \, \bm y} \right) 
  + \frac{1}{2 \, N_c \, C_F} \, {\hat O}_{\bm u \, \bm y} \right\rangle \; .
 \end{eqnarray}
\end{subequations}

The spin-dependent and spin-averaged cross sections $d(\Delta \sigma)$
and $d\sigma_{unp}$ for quark production from \eqref{eq-Defn STSA}
read
\begin{subequations}
 \label{eq-observables}
 \begin{eqnarray}
   d(\Delta\sigma^{(q)}) = \frac{C_F}{(2\pi)^3} \, \frac{\alpha}{1-\alpha}  \, 
   \int d^2 x \, d^2 y \, d^2 z
   \, e^{-i \bm k \cdot (\bm z - \bm y)} \notag \\ 
   \times \, \Phi_{pol} (\bm z - \bm x \, , \, \bm y - \bm x, \alpha) \
   \mathcal{I}^{(q)}_{anti} (\bm x \, , \, \bm y \, , \, \bm z) \label{dsigmaq} \\
   d\sigma^{(q)}_{unp} = \frac{C_F}{2 \, (2\pi)^3} \, \frac{\alpha}{1-\alpha} \, 
   \int d^2 x \, d^2 y \, d^2 z
   \, e^{-i \bm k \cdot (\bm z - \bm y)}  \notag \\ 
   \times \, \Phi_{unp} (\bm z - \bm x \, , \, \bm y - \bm x, \alpha) \
   \mathcal{I}^{(q)}_{symm} (\bm x \, , \, \bm y \, , \, \bm z) \; , \label{dsigmaq_unp}
 \end{eqnarray}
\end{subequations}
where the wave functions squared are given by Eqs.
\eqref{eq-unpolarized wavefn}, \eqref{eq-polarized wavefn}, and the
interactions are given by Eqs.~\eqref{eq-symmetrized interaction}.
Eqs.~\eqref{eq-observables} are the main results of this work: when
substituted into \eq{eq-Defn STSA} they give the single-transverse
spin asymmetry $A_N$ generated in quark production by the $C$- and
$T$-odd interactions with the target.



\section{Properties of the Obtained STSA}

The expressions in Eqs.~\eqref{eq-observables} are hard to evaluate in
general. To understand the main properties of the STSA contribution
resulting from our mechanism, we used the amplitudes ${S}_{\bm x \,
  \bm y}$ and ${O}_{\bm x \, \bm y}$ evaluated in the Glauber-Mueller
[\refcite{Mueller:1989st}] multiple-rescattering approximation and
determined the corresponding STSA in the large-$N_c$ limit after
making several crude approximations described in
[\refcite{Kovchegov:2012ga}]. The results are summarized below.

\subsection{Transverse Momentum and Feynman-$x$ Dependence}

First of all, in the large-$k_T$ limit the STSA due to our mechanism
is (assuming that the unpolarized cross section $d\sigma_{unp}$ is
dominated by gluon production)
\begin{eqnarray}
  A_N^{(q)} \bigg|_{k_T \gg Q_s} \propto \frac{k^2}{k_T^{6}} \propto 
  \frac{1}{k_T^5} \label{dsigmaq7} 
\end{eqnarray}
which is a steeply-falling function of $k_T$. This indicates that in
the standard factorization framework our STSA generating mechanism
originates in some higher-twist operator.

\begin{figure}[h]
  \begin{center}
     \includegraphics[width=0.6 \textwidth]{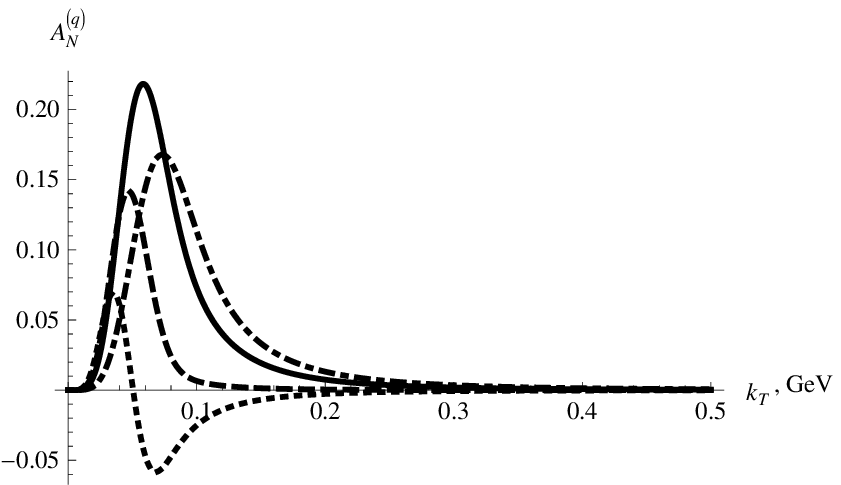} \\ \vspace*{3mm} \includegraphics[width=0.6 \textwidth]{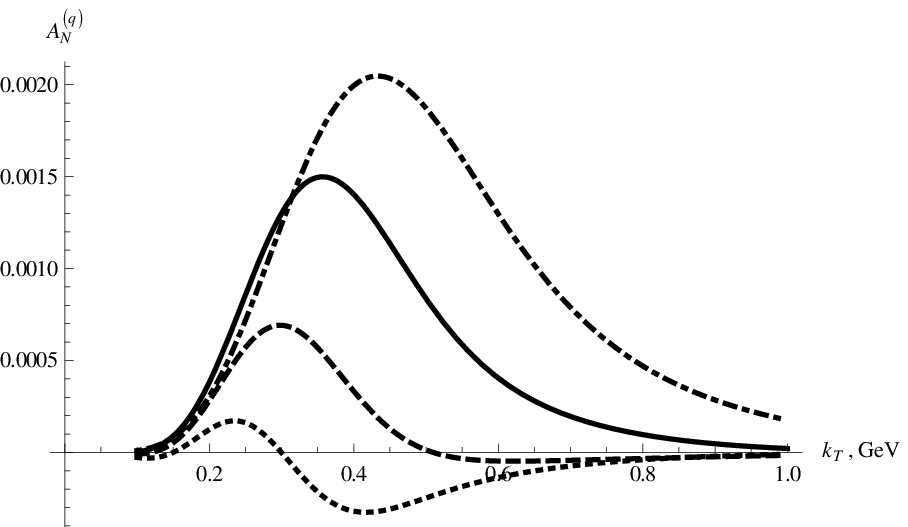}
  \end{center}
  \caption{Quark STSA in our mechanism for the proton target plotted
    as a function of $k_T$ for different values of the longitudinal
    momentum fraction $\alpha$ carried by the produced quark: $\alpha
    = 0.9$ (dash-dotted curve), $\alpha = 0.7$ (solid curve), $\alpha
    = 0.6$ (dashed curve), and $\alpha = 0.5$ (dotted curve). The
    infrared cutoffs are 2.1 fm in the top panel and 1.3 fm in the
    bottom panel.}
\label{AN} 
\end{figure}

The $k_T$-dependence of $A_N$ is illustrated over a broader momentum
range in \fig{AN}. Note that the asymmetry is a non-monotonic function
of $k_T$ which even changes sign (it has zeros/nodes). This seems to
be in qualitative agreement with some of the experimental data on
STSA. The maximum of $A_N$ in the saturation framework is correlated
with the saturation scale. Another important feature is that the
asymmetry in \fig{AN} increases with the increasing momentum fraction
$\alpha$ (an analogue of the Feynman-$x$ variable) for most values of
$\alpha$: this is also in qualitative agreement with experiment.

An important feature of our STSA is that it is proportional to
\begin{equation}
  \label{eq:grad}
  A_N^{(q)} \sim \int d^2 b \, [{\bm \nabla} T ({\bm b})]^2 
\end{equation}
with $T ({\bm b})$ the nuclear profile function. Since the variation
of $T ({\bm b})$ is largest in the peripheral collisions, our STSA
seems to be dominated by large impact parameter scattering and is
sensitive to the edges of the target. This explains the difference
between the two panels in \fig{AN}, which are different in the
infrared cutoff used to limit the large-$b$ integration. A full
analysis of the obtained formula without crude approximations made in
[\refcite{Kovchegov:2012ga}] should determine whether the saturation
scale is large enough in the peripheral collisions dominating STSA for
the perturbative approach to be valid.


\subsection{Atomic Number and Centrality Dependence}

\begin{figure}[b]
\begin{center}
  \includegraphics[width=0.6 \textwidth]{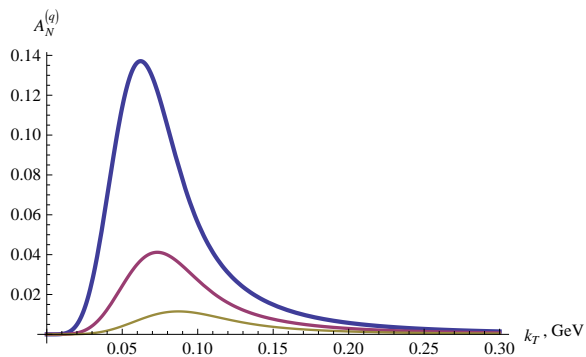}
\end{center}
\caption{Quark STSA in our mechanism plotted as a function of $k_T$
  for different values of the target radius: $R = 1$~fm (top curve),
  $R = 1.4$~fm (middle curve), and $R = 2$~fm (bottom curve) for
  $\alpha = 0.7$.}
\label{ANrad} 
\end{figure}

The dependence of our STSA on the size of the target for $k_T \approx
Q_s$ is
\begin{equation}
  \label{Adep}
  A_N^{(q)} (k_T \approx Q_s) \sim \frac{1}{Q_s^7} \sim A^{-7/6},
\end{equation}
if $Q_s^2 \sim A^{1/3}$. This is a very steep falloff of STSA with the
atomic number of the nuclear target, which is also illustrated in
\fig{ANrad}, where $A_N$ is plotted for several target radii. We thus
predict that, in our formalism, STSA should be much smaller in the
polarized proton--nucleus collisions than in the polarized
proton--proton collisions.


\subsection{STSA in Photon Production}
\label{subsec-Photon STSA (est)} 

The mechanism we employed allows one to calculate the STSA of prompt
photons as well. The calculation yields [\refcite{Kovchegov:2012ga}]
\begin{eqnarray}
  \label{dsigmaph1}
  d(\Delta\sigma^{(\gamma)}) = \frac{i}{(2\pi)^3} \,  
  \int d^2 x \, d^2 y \, d^2 z
  \, e^{-i \bm k \cdot (\bm z - \bm y)} \, \Phi_{pol} (- \bm z \, , \, - \bm y, \alpha) \notag \\ \times \,
  \left[ O_{{\bm x} + (1-\alpha) \, {\bm z},  \, {\bm x} + (1-\alpha) \, {\bm y}} 
    - O_{{\bm x} , \, {\bm x} + (1-\alpha) \, {\bm y}} 
    - O_{{\bm x} + (1-\alpha) \, {\bm z}, \, {\bm x}} \right]
\end{eqnarray}
where $\Phi_{pol}$ is given by \eq{eq-unpolarized wavefn} with the
$\as \to \alpha_{EM} \, Z_f^2$ replacement. This expression is zero
since
\begin{equation}
  \label{zero2}
  \int d^2 x \, 
   \,  O_{{\tilde {\bm z}} + {\bm x}, \, {\tilde {\bm y}} + {\bm x}} =0
\end{equation}
due to the fact that the odderon amplitude is an anti-symmetric
function of its transverse coordinate arguments,
\begin{equation}
  \label{Oanti}
  O_{\bm x \, \bm y} = - O_{\bm y \, \bm x}.
\end{equation}

We thus have an exact result that in our mechanism the photon STSA is zero,
\begin{equation}
  \label{dsigmaph2}
  A_N^{(\gamma)} = 0,
\end{equation}
at least in the order of approximation considered here.


\section{Outlook}

\begin{figure}[h]
\begin{center}
  \includegraphics[width=\textwidth]{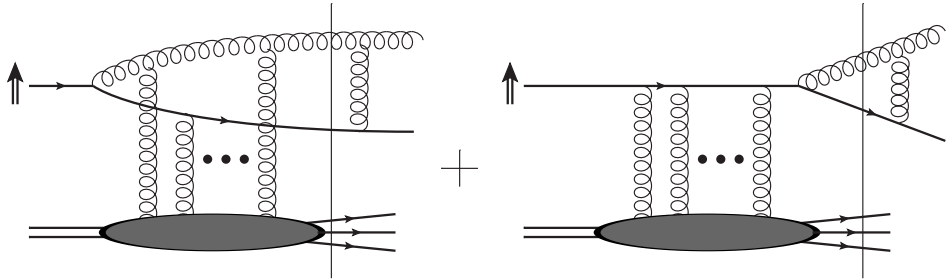}
\end{center}
 \caption{Diagrams incorporating the Sivers effect into our formalism.}
\label{Sivers} 
\end{figure}

In the future it will be important to explore other ways of generating
STSA in the saturation/CGC formalism. Above we studied STSA
originating in the interaction with the unpolarized target. However,
as we mentioned, it may also originate in the wave function
squared. This would be the Sivers mechanism as studied in
[\refcite{Brodsky:2002cx}]. The analogue of this effect is illustrated
in \fig{Sivers}, where the STSA should be obtained by the interference
between the diagrams in Figs.~\ref{Sivers} and \ref{qtoqGampl}.
The relative phase would arise due to the extra rescattering with the
polarized projectile, and is depicted by the cut giving the imaginary
part. (Indeed \fig{Sivers} shows only one of the relevant diagrams.)
The analysis of this contribution is left for future work
[\refcite{KS2}].


\section*{Acknowledgments}

This research is sponsored in part by the U.S. Department of Energy
under Grant No. DE-SC0004286.



\providecommand{\href}[2]{#2}\begingroup\raggedright\endgroup


\end{document}